\documentclass{ws-procs975x65}

\begin{document}

\title{CONSTRAINING THE CENTRAL MAGNETIC FIELD OF MAGNETARS}

\author{BANIBRATA MUKHOPADHYAY$^1$,~~MONIKA SINHA$^{2,3}$}

\address{1. Department of Physics, Indian Institute of Science,
Bangalore, 560012,\\
E-mail: bm@physics.iisc.ernet.in\\
2. Institute for Theoretical Physics, J. W. Goethe-University, D-60438 Frankfurt am Main\\
3. Indian Institute of Technology Rajasthan, Old Residency Road, Ratanada, Jodhpur 342011,
E-mail: ms@iitj.ac.in
}

\begin{abstract}
The magnetars are believed to be highly magnetized neutron 
stars having surface magnetic field $10^{14} - 10^{15}$ G. 
It is believed that at the center, the magnetic field may be 
higher than that at the surface. We study the effect of the 
magnetic field on the neutron star matter.
We model the nuclear 
matter with the relativistic mean field approach considering 
the possibility of appearance of hyperons at higher density. 
We find that the effect of magnetic field on the matter of 
neutron stars and hence on the mass-radius relation is important, 
when the central magnetic field is atleast of the order of 
$10^{17}$ G. Very importantly, the effect of strong magnetic 
field reveals anisotropy to the system. Moreover, if the central 
field approaches $10^{19}$ G, then the matter becomes 
unstable which limits the maximum magnetic field at the center 
of magnetars.

\end{abstract}


\bodymatter

\section{Introduction}\label{sec1}
Anomalous X-ray pulsars and soft $\gamma$-ray repeaters are
observationally identified with highly magnetized neutron stars,
known as magnetars,
with surface magnetic field $\sim 10^{14}-10^{15}$ G
\cite{natur}.
The processes of supernova collapse will leave behind a strongly
non-uniform frozen-in field distribution. Also any dynamo
mechanism generating fields will carry the imprint of inhomogeneous
density profile in the star. Thus, to maintain the local magneto-static 
equilibrium, more realistic treatment of the equation of state (EoS)
of matter for a magnetar requires inclusion of gradually increasing magnetic 
field from surface to center.
Massive compact stars are likely to develop exotic cores with
one possibility of appearance of hyperons with the increasing 
density.
In the present work, considering the radial profile of the magnetic
field and carefully analyzing the different components of the
field, we show that the pressure of the magnetar matter parallel to the 
magnetic field exhibits instability.

\section{Model of magnetar matter}
To construct the model of dense matter, we employ non-linear 
Walecka mean field theory \cite{walecka,bb} of nuclear matter including
the possibility of appearance of hyperons and muons at higher
density.
In the presence of magnetic field, the Lagrangian density 
of the system is
${\cal L} = \sum_b{\cal L}_b +\sum_l{\cal L}_l  - \frac14 F_{\mu \nu}F^{\mu \nu}$,
with ${\cal L}_b$  and ${\cal L}_l$ are the baryonic and leptonic  
Lagrangian densities respectively in the presence of magnetic
field \cite{yz,czl} and $F_{\mu\nu}$ the electro-magnetic field 
tensor. For details, see Ref.~\refcite{sms}.

Total energy density and pressure of the system can be obtained
by considering the energy-momentum tensor of the system
$T^{\mu \nu} = T^{\mu \nu}_m + T^{\mu \nu}_f$,
where $T^{\mu \nu}_m$ and $T^{\mu \nu}_f$ are the matter 
and field parts respectively. In the absence of electric 
field, 
and in the rest frame of fluid
\begin{equation}
T^{\mu\nu} = \left[\begin{array}{cccc} 
                    \varepsilon_m+ \frac{B^2}{8\pi} & 0 & 0 & 0\\
                    0 & P-M B+ \frac{B^2}{8\pi} & 0 & 0\\
                    0 & 0 & P-M B+ \frac{B^2}{8\pi} & 0\\
                    0 & 0 & 0 & P- \frac{B^2}{8\pi} 
               \end{array}\right],
\label{tmunu}
\end{equation}
when the magnetic field is considered to be along $z$-direction 
with  $B^\mu B_\mu = -  B^2$, $B$ is
the magnitude of magnetic field, $M$ the magnetization
per unit volume, $P$ and $\varepsilon_m$ are respectively the pressure and energy density of the
matter. This clearly shows anisotropic nature of the pressure in the presence 
of (strong) magnetic field. $\varepsilon_m$ is
calculated using the charged single particle energy 
$E_n =  \sqrt{p_z^2 + m^2 + 2 n e |Q| B}$,
considering the quantized phase space in the presence of magnetic
field, where $p_z$ is the component of momentum along $z$-axis, $m$ the mass, 
$e|Q|$ the total charge, with $e$ being the electron's charge, of the particle,
$n$ is the number of occupied Landau level.
Then matter pressure is 
$P=\sum_b \mu_b n_b + \sum_l \mu_l n_l - \varepsilon_m$, where
$\mu_{b,l}$ and $n_{b,l}$ are respectively the chemical potentials and 
number densities for baryons ($b$) and leptons ($l$). The density profile of 
the magnetic field is modeled as \cite{prl}
\begin{equation}
\label{profile}
B\left(\frac{n_b}{n_0}\right)= B_s+B_c\left\{1-\exp\left[{-\beta \left(
\frac {n_b}{n_0} \right)^\gamma}\right]\right\},
\end{equation}
where $\beta$ and $\gamma$ are two parameters, $n_b$ and $n_0$ are respectively the
number densities of matter and nuclear matter, $B_s$ and $B_c$ are 
respectively the magnitudes of the magnetic field in the surface and center of the underlying magnetar.

\section{Results}\label{discussion}

\begin{figure*}
\begin{center}
\includegraphics[angle=0,width=12cm,height=5.8cm]{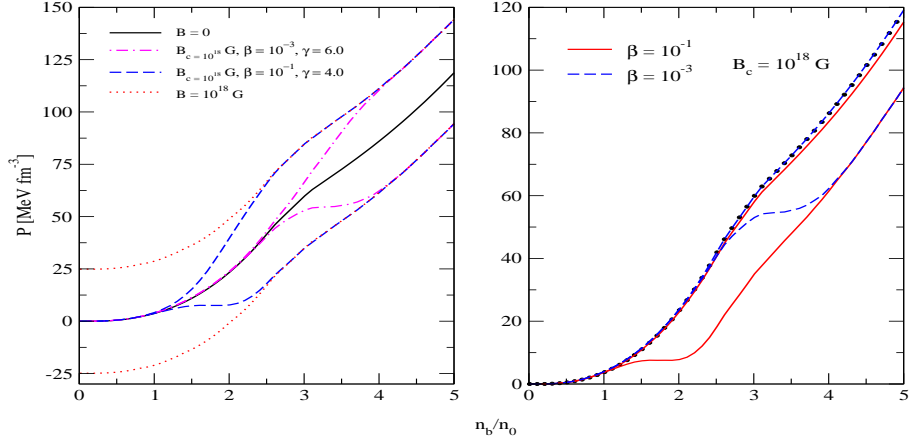}
\caption{
Left panel (a):
Variation of total pressure as a function of normalized 
baryon number density for fixed fields $B = 0$ (solid line), and 
$B_c = 10^{18}$~G and several density profiles: $\beta = 10^{-3}$,
$\gamma = 6$ (dot-dashed lines), $\beta = 10^{-1}$,
$\gamma = 4$  (dashed lines), and $\beta \to \infty$,
i.e,  $B =$ constant (dotted lines).
For each pair of curves the upper branch is for $P_\perp$ and the lower branch
for $P_\parallel$.  
Right panel (b):
Variation of total pressure as a function of normalized baryon number density for 
different magnetic field profiles and $B_c = 10^{18}$ G. The dots show the 
reference case $B = 0$. The solid and dashed lines correspond to $\beta = 0.1$
and $0.001$ respectively. For each $\beta$ we choose a pair of $\gamma$s; in the first 
case we have $\gamma = 1$ (upper), $4$ (lower), whereas in the
second case $\gamma = 1$ (upper), $6$ (lower). 
}
\label{eos1}
\end{center}
\end{figure*}

Figure \ref{eos1}a shows EoS for hypernuclear matter 
in strong and fixed magnetic field profiles. 
For non-zero magnetic field, the pressure splits into the parallel ($P_\parallel$) and transverse 
($P_\perp$) components,
and exhibits anisotropy. 
It is seen that the low-density behavior of EoS with constant magnetic 
field implies unrealistically large anisotropic
magnetic field up to the surface of the star, which is inconsistent with the inferred surface
magnetic field ($\sim 10^{15}$ G) of magnetars.
It is seen from Fig. \ref{eos1}b that
for a given value of $\beta$, the EoS becomes softer with increasing $\gamma$. Consequently,
beyond a certain critical $\gamma$ and in a certain density regime, $P$ ceases to increase 
(and eventually decreases) with the further increase in $n_b$. 
This implies the onset of instability of matter above that value of density for that particular 
$B_c$ and magnetic field profile. 
We also show the results for each $\beta$ with $\gamma=1$ (minimum value). Note
that the maximum  $\gamma$ is taken in such a way that $P$ forms a plateau as a function $n_b$. 
Furthermore, it is evident that with the decrease of $\beta$, the instability occurs at 
larger values of $\gamma$ and $n_b$.

The instability arises due to the negative contribution from the field energy density 
to the pressure of magnetized baryons and leptons in the direction of the magnetic field, which
is evident from Eq. (\ref{tmunu}). With the increase of $n_b$, more negative contribution is added to 
$P$, and consequently at a certain $n_b$, $P$ ceases to increase and 
then decreases with the increase of $n_b$, rendering instability. 

\section{Conclusion}

We have found that for sufficiently large magnetic fields with $B_c\sim 10^{18}$ G, the magnetar
matter becomes
unstable. The instability is associated with the anisotropic effects arised due to
the magnetic field. The onset of instability depends on
the magnetic field profile and $B_c$, which puts a natural upper bound for the central magnetic field 
of neutron stars, which is $5\times 10^{18}$ G. \\

This work was partially supported by the grant ISRO/RES/2/367/10-11 (B.M.)
and the Alexander von Humboldt Foundation (M.S.).



\end{document}